\documentclass[useAMS,usenatbib]{mn2e}
\usepackage{graphicx}
\usepackage{amssymb}

\newcommand{\mnras}{{MNRAS}}
\newcommand{\apj}{{ApJ}}
\newcommand{\aap}{{A\&A}}
\newcommand{\actaa}{{AcA}}
\newcommand{\pasj}{{PASJ}}
\newcommand{\pasp}{{PASP}}
\newcommand{\apjl}{{ApJL}}
\newcommand{\apjs}{{ApJS}}
\newcommand{\aj}{{AJ}}
\newcommand{\nat}{{Nature}}
\newcommand{\apss}{{Ap\&SS}}

\title[On rates of lensed supernovae]{On rates of supernovae strongly lensed by galactic haloes in Millennium Simulation}

\author[Zuzanna Kostrzewa-Rutkowska,{\L}ukasz Wyrzykowski and Micha{\l} Jaroszy{\'n}ski]{Zuzanna Kostrzewa-Rutkowska$^{1}$\thanks{E-mail: zkostrzewa@astrouw.edu.pl}, {\L}ukasz Wyrzykowski$^{1,2}$ and Micha{\l} Jaroszy{\'n}ski$^{1}$\\
$^{1}$Warsaw University Astronomical Observatory, Al. Ujazdowskie 4, 00-478 Warszawa, Poland\\
$^{2}$Institute of Astronomy, University of Cambridge, Madingley Road, Cambridge CB3 0HA, UK}

\begin{document}

\date{2012 Oct 31}

\pagerange{\pageref{firstpage}--\pageref{lastpage}} \pubyear{2012}

\maketitle

\label{firstpage}

\begin{abstract}
We make use of publicly available results from N-body Millennium Simulation to create mock samples of lensed supernovae type Ia and core-collapse.
Simulating galaxy-galaxy lensing we derive the rates of lensed supernovae and find than at redshifts higher that 0.5 about 0.06 per cent of supernovae will be lensed by a factor two or more. 
Future wide field surveys like Gaia or LSST should be able to detect lensed supernovae in their unbiased sky monitoring. 
Gaia (from 2013) will detect at least 2 cases whereas LSST (from 2018) will see more than 500 a year. 
Large number of future lensed supernovae will allow to verify results of cosmological simulations. 
The strong galaxy-galaxy lensing gives an opportunity to reach high-redshift supernovae type Ia and extend the Hubble diagram sample. 
\end{abstract}


\section{Introduction}
Gravitational lensing is a successful tool allowing observing surveys to reach to potentially smaller, fainter, and less massive objects as it does not rely on extremely deep imaging. 
In strong lensing a massive foreground galaxy deflects the light from a background object, resulting in multiple images of the source being seen. 
In addition the source is typically magnified by a factor of $\sim$10 (e.g. \citealt{2011ApJ...734..104N}).
This fact allows for studying of objects which are fainter and smaller than otherwise possible (\citealt*{1992grle.book.....S}; \citealt{2006glsw.conf....1S}).
We currently know a few hundred of strong lens systems.
Many such system have been discovered in the radio-domain (e.g. \citealt{2000AJ....120.2868W}; \citealt{2003MNRAS.341...13B}; \citealt{2003MNRAS.341....1M}), through spectroscopic or colour selection from different surveys, e.g. \cite{2006ApJ...638..703B}, \cite{2008ApJ...682..964B}, \cite{2006ApJ...640..662T} (SLACS), \cite{2009MNRAS.392..104B} (CASSOWARY), \cite{2012ApJ...744...41B} (BELLS),  \cite{2009A&A...502..445L}, \cite{2012arXiv1202.3852G} (SL2S), \cite{2011MNRAS.417.1601T} (SWELLS), \cite{2010Sci...330..800N} (HerschelATLAS) and in Hubble Space Telescope COSMOS survey \citep{2008ApJS..176...19F} at optical wavelengths.
We observe quasars, radio sources and various types of galaxies as source population, which provide a wide range of applications of the lensing phenomenon.
Strong lenses have been used, for example, to investigate the density profile of lensing galaxies, the evolution of galaxies and the properties of dark energy (e.g. \citealt{2008ApJ...682..964B}; \citealt{2010ApJ...721L.163A}; \citealt{2011ApJ...734..104N}; \citealt{2011ApJ...727...96R}; \citealt{2011ApJ...737..102S}; \citealt{2012ApJ...757...82B}; \citealt{2012ApJ...750...10S}).

Supernovae (SNe) are divided into two main types I and II based on their spectra, with a secondary distinction based on light-curve shape. 
Supernovae type Ia (SNe Ia) are widely used as a cosmological tool to analyse the Hubble diagram and to constrain parameters $\Omega_\mathrm{M}, \Omega_{\Lambda}, w$   (\citealt{1998AJ....116.1009R}; \citealt{2006A&A...447...31A}; \citealt{2011ApJS..192....1C}; \citealt{2011ApJ...737..102S}). 
SNe Ia remain, at present, one of the most direct and mature method of probing the dark energy.
Thought to be the result of the thermonuclear explosion of a white dwarf star exceeding the Chandrasekhar mass limit by accreting mass from its companion star in a binary system, or by merging with another white dwarf, they are standardisable candles which explode with almost the same brightness everywhere in the Universe due to the uniformity of the triggering mass and hence the available nuclear fuel. 
Their cosmological use takes advantage of simple empirical relations between their luminosity and other parameters.
The application of relations between SN Ia light-curve-shape, colour, and host galaxy properties provides robust distance estimates which allow SNe Ia to be used to measure cosmological parameters.
The SNe Ia samples are now sufficiently numerous that we are able to measure dark energy (e.g. \citealt{2011ApJ...737..102S})

The core-collapse supernovae (SNe CC) are thought to be the final evolutionary phase of massive stars with initial mass larger than 8 M$_{\sun}$.
The progenitors of CC supernovae are massive stars, either single or in binary systems, that complete exothermic nuclear burning, up to the development of an iron core that cannot be supported by any further nuclear fusion reactions or by electron degeneracy pressure. 
Due to the brief lifetime of their progenitor stars (shorter than the typical cosmological time scale), the rate of SNe CC occurrence closely follows the current star formation rate in a stellar system (e.g. \citealt{2012A&A...537A.132B}). 
SNe CC are responsible for a significant fraction of heavy elements present in the Universe and their energetic impacts perhaps trigger further star formation.
They are also relevant to many astrophysical issues, for example, associated with the physical and chemical evolution of the galaxies and the local environment, the production of neutrinos, cosmic rays and gravitational waves.
Moreover, SNe CC seem to be particularly promising to measure cosmological distance, in addition to type Ia SNe according to \cite{2011arXiv1109.1781P}.

Forasmuch as the light collection capabilities of telescopes are limited, our prospect to study high-redshift supernovae is strongly tentative.
The high-redshift SNe Ia are necessary to explore cosmological parameters more accurately.
The strong lensing acting as a cosmic telescope can provide an extra insight into higher redshift supernovae and open a new area of this research (e.g. \citealt{2002A&A...393...25G}; \citealt{2010MNRAS.405.2579O}).

In this paper, we present predictions for expected numbers of supernovae type Ia and core-collapse which are gravitationally lensed in the most common galaxy-galaxy lensing.
We investigate how many lensed supernovae might be detected in currently on-going and future large-scale surveys like Gaia or LSST.

For many of the existing and planned surveys exploring the transient sky detecting supernovae is one of the main scientific goals.
The Supernovae Search in Stripe 82 of SDSS programme yielded about 500 detections of Supernovae and contains almost 4 million stellar objects and galaxies monitored in the $u$, $g$, $r$, $i$, and $z$ bands (\citealt{2008MNRAS.386..887B}; \citealt{2010ApJ...724..502H}).

ESA's Gaia mission will collect astrometric, photometric, spectro-photometric and spectroscopic measurements for one billion celestial objects from the entire sky down to $V\sim20$ mag.
The mission is designed to last 5 years, with a possible one-year extension, after a launch currently planned for September 2013 \citep{2012IAUS..285..425W}.
Gaia is expected to detect about 6300 SNe (to 19 mag), of which 85 per cent will be Ias, of which 30 per cent will be detected before maximum (\citealt{2003MNRAS.341..569B}; \citealt{2012Ap&SS.tmp...66A}). 
The maximum redshift detectable by Gaia for SN type Ia is about 0.15, however availability of low-resolution spectra for all objects will significantly decrease the rate of false-alerts and will allow for immediate supernovae typing and redshift estimation (\citealt{2012arXiv1210.5007W}; Blagorodnova et al. in prep.).

The most ambitious large-scale survey currently planned in the optical domain is the Large Synoptic Survey Telescope (LSST) (\citealt{2009arXiv0912.0201L}). 
LSST will be a wide-field ground-based system designed to obtain multiple filter images covering the sky visible from Northern Chile. 
The current baseline design, with an 8.4m (6.7m effective) primary mirror, a 9.6 $\mathrm{deg^2}$ field of view and a 3.2 Gigapixel camera, will allow for about 20,000 square degrees of sky in total to be covered.
One of the main scientific goals of LSST is exploring transient optical sky (\citealt{2008arXiv0805.2366I}).

This paper is organised as follows. 
In Section 2 we describe the Millennium Simulation with its applications. 
We then present supernovae rates used in our simulations in Section 3 and in Section 4 we describe lensed supernovae simulation and show our results.
After that we discuss the results and conclude in Section 5. 
Throughout this paper we assume a flat $\Lambda$-cold-dark-matter ($\Lambda$CDM) cosmological model with parameters $\Omega_\Lambda=0.73$, $\Omega_\mathrm{M}=0.27$ and $H_0=71 \mathrm{km~s^{-1} Mpc^{-1}}$, $h=0.71$ as favoured by the seven year Wilkinson Microwave Anisotropy Probe (WMAP) results (\citealt{2011ApJS..192...18K}).

\section{Millennium Simulation and its applications}
In this study we rely on the Millennium Simulation \citep{2005Natur.435..629S},\footnote{In the Millennium Simulation the $\Lambda$CDM cosmological model with parameters obtained by WMAP 1-year-observations ($\Omega_\Lambda=0.75$, $\Omega_\mathrm{M}=0.25$ and $H_0=73~\mathrm{km~s^{-1} Mpc^{-1}}$) was assumed. 
For this kind of simulation (specifically the analysis of gravitational instability) two parameters play an important role: the amplitude of density fluctuations, $\sigma_8=0.9$ (as compared with the current value of 0.8), and the density perturbation spectral index $n=1$ (0.963 respectively). 
The higher value of parameter $\sigma_8$ in the simulation implies that the gravitationally bound structures in the mechanical equilibrium appear earlier there than they did in reality.
The discrepancies in the evolution between the simulation and the Universe increase with the object mass. 
We are not able to transform the results from the simulation to comply with todays knowledge. 
We are aware that the lensing probability may be overestimated, although for supernovae bursting at redshifts of less than 2.5 (like in our study) the effect should not be dramatic.
} which offers high spatial and time resolution within a large cosmological volume, ideal for studying lensing at various scales. The Simulation describes the evolution of matter distribution and formation of the cosmic structures in a cube of co-moving size $500 h^{-1}\mathrm{Mpc}$ and extends the possibility of realistic modelling of the light propagation in the Universe.
In particular the matter distribution is described by the positions of the $\sim 10^{10}$ simulation particles known at 64 epochs representing redshift from 0 to 127. 
The positions, virial masses, virial radii, and virial velocities of gravitationally bound haloes (containing at least 20 simulation particles) obtained by \citet{2007MNRAS.375....2D}, \citet{2007MNRAS.379.1143B} are taken from the Virgo - Millennium Database website.\footnote{
http://gavo.mpa-garching.mpg.de/Millennium/
} 
In the whole study we consider the haloes as the singular isothermal ellipsoids (SIE) (\citealt{1994A&A...284..285K}, \citealt{1997ApJ...487...42K}).  
Every halo from the simulation is a probable lens which will be described by a few parameters: the Einstein radius, the axis ratio, the position angle. 
The Einstein radius depends on the mass of the dark matter halo (or here on the virial velocity of the halo), the distance between the observer and the source and the distance between the lens and the source: 
\begin{eqnarray}
\Theta_{\mathrm{SIE}}=4\pi\frac{\sigma^2}{c^2}\frac{D_{ls}}{D_{os}},
\end{eqnarray}
where $\sigma$ is the velocity dispersion of the halo, $c$ is the light speed, $q$ is the axis ratio, $D_{ls}, D_{os}$ are cosmological angular diameter distances from the lens to the source and from the observer to the source, respectively.
The haloes in the simulation are described by the virial velocity which satisfies the relation for rotation velocity: $v_{vir}=\sqrt{\frac{GM_{vir}}{R_{vir}}}$, where $M_{vir}$ is the virial mass, $R_{vir}$ is the virial radius, $G$ is the gravitational constant.
Hence, by taking into account the relation $v^2_{rotation}=2 \sigma^2$, the Einstein radius in our study can be expressed by: 
\begin{eqnarray}
\Theta_{\mathrm{SIE}}=2\pi\frac{v_{vir}^2}{c^2}\frac{D_{ls}}{D_{os}}.
\end{eqnarray}

We, therefore, use the results from the Millennium Simulation database to obtain the distributions of feasible lens properties and the lensing probability.
 
\subsection{The lens velocity distribution}
\begin{figure}
\includegraphics{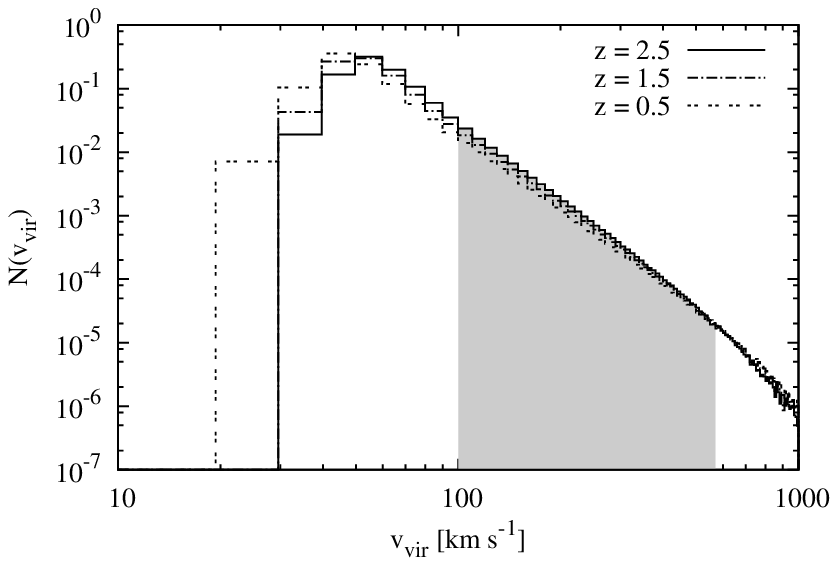}
\includegraphics{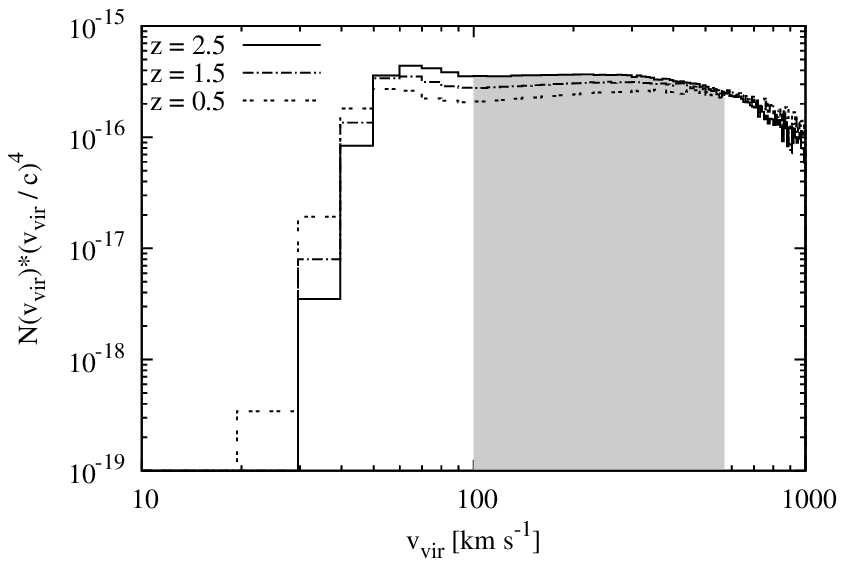}
\caption{Upper panel: The normalized distribution of the virial velocity of dark matter haloes in the Millennium Simulation.
Lower panel: The contribution of each velocity bin to lensing probability.
Shaded region - the range of typical galaxy virial velocities.}
\label{figvvir}
\end{figure}

For each redshift epoch we obtain the virial velocity distribution of gravitationally bound haloes versus redshift which we use to randomise the Einstein radius.
As it can be seen in Figure \ref{figvvir} (upper panel) that the velocities of the haloes extend from 20 to 1000 $\mathrm{km~s^{-1}}$. 
In this study we only investigate lensing by individual galaxies, as it is far more common than galaxy cluster lensing. 
Indeed, there are many advantages of observing massive galaxy clusters to find lensed supernovae due to their large regions of high magnification. 
These aspects were already fully discussed in \cite{2009A&A...507...71G} and  \cite{2011A&A...536A..94R}.
We assume that the velocity dispersion of a typical galaxy should be in range between 70 and 400 $\mathrm{km~s^{-1}}$ ($100 < v_{vir} < 570~\mathrm{km~s^{-1}}$), following the fitting of a modified Schechter function to the observed SDSS sample (\citealt{2007ApJ...658..884C}).
This restricts us from using very massive haloes (which can be considered as galaxy clusters).
Moreover, we neglect a group of low-mass haloes (expected as dwarf galaxies), as on these scales ($v_{vir} < 100~\mathrm{km~s^{-1}}$) cold dark matter simulations appear to overestimate the amount of structure, to a degree that appears inconsistent with current observations (e.g. \citealt{2008MNRAS.391.1685S}). 
We cannot accurately take into account the effect of dwarf haloes on lensing probability, because there is insufficient data on their real distribution. 
We therefore exclude them altogether from this study. 
However, one can estimate that their inclusion would increase the lensing probability by 6-10 per cent.

\subsection{Galaxy-galaxy lensing probability}
\begin{figure}
\includegraphics{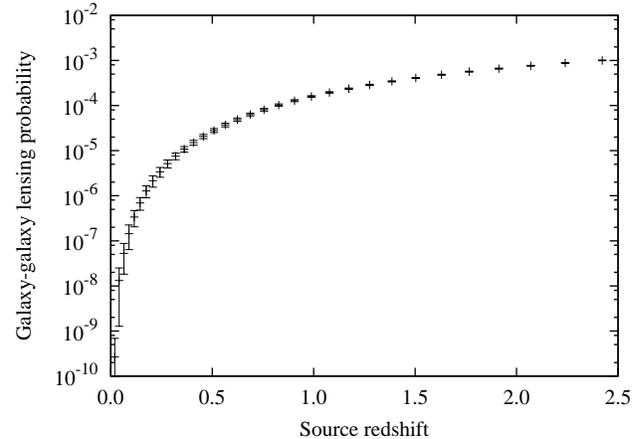}
\caption{The whole-sky averaged galaxy-galaxy lensing probability as a function of source redshift.}
\label{figlike}
\end{figure}

In constructing the matter layers we follow \cite{2008MNRAS.388.1618C}, randomly rotating and shifting simulation cubes corresponding to different epochs, which eliminates the consequences of periodic boundary conditions used in their calculations.
If we consider a source at epoch $n$, then there are $(n-1)$ layers between the observer and the source, where the lens can be situated. 
We select ten independent directions on the sky and examine fields covering $3^{\circ}\times3^{\circ}$ each. 
Each halo from the cone contributes to the strong lensing probability as $\pi \Theta_{\mathrm{SIE}}^2/(3^{\circ}\times3^{\circ})$, see the lower panel of Figure \ref{figvvir}. 
Afterwards, we average the probabilities from each of the ten fields to estimate the probability of strong lensing on the whole sky. 
As a result we obtain the strong lensing probability for each source epoch, represented by the epoch redshift, see Figure \ref{figlike}. 
Lensing probability rises sharply in the redshift regime from 0 to 0.5, reaching $10^{-5}$ and then flattens at about $10^{-4}$ to $10^{-3}$ for higher redshifts. 
We acknowledge the fact that by treating each halo as a SIE we add extra mass to the line of sight.
That leads to an overestimation of the Universe density and therefore of the magnification and lensing rates.
We examined a few hundred haloes randomly chosen from the Millennium Simulation compensated by a constant, negative surface matter density disk (corresponding to the convergence $\kappa$) and we claim that the cross-section for multiple images does not change in such a model.
The separation of images changes by a factor $1/(1+|\kappa|)$ and the amplification by $1/(1+|\kappa|)^2$, where the value of parameter $|\kappa|$ is really small ($<10^{-3}$), hence we conclude the problem of overdensity can be neglected in this study.

\subsection{The lens location probability}
\begin{figure}
\includegraphics{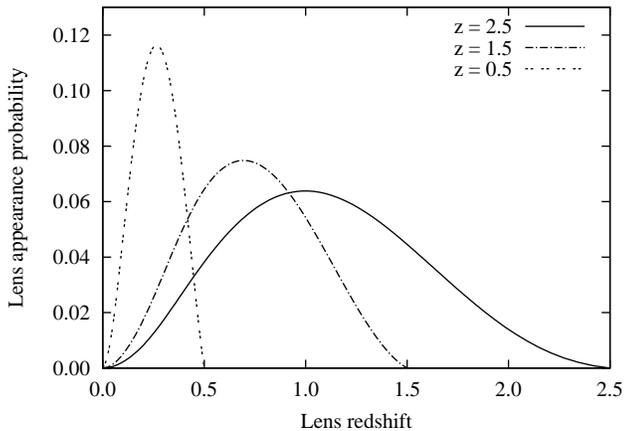}
\caption{The lens location probability in the Millennium Simulation for sources at different redshifts.}
\label{figlens}
\end{figure}

After we choose the source redshift, we are interested in the likelihood of a lens presence in every redshift bin between the observer and the source. 
For every source epoch we saved the contribution of every redshift layer to the total probability, so we are able to randomise the lens redshift with this distribution.
Figure \ref{figlens} shows the lens location likelihood for sources at three different redshifts.

\subsection{The shear distribution}
\begin{figure}
\includegraphics{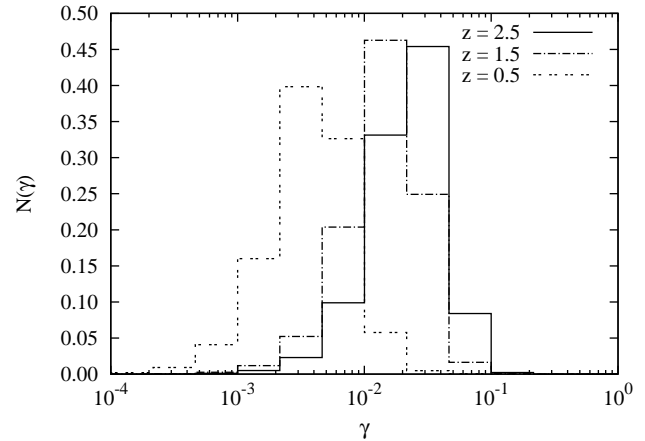}
\caption{The shear distribution in the Millennium Simulation for the sources at different redshift.}
\label{figshear}
\end{figure}

The smoothed distribution of matter and the positions and properties of gravitationally bound haloes are sufficient to describe the light propagation in terms of convergence. 
In \citet{2010AcA....60...41J} use of the Born approximation enabled the construction of the mass distribution and the calculation of the convergence and shear maps.
We utilise those results to obtain the distribution of the shear, which will contribute to the lensing signal as an effect of the lens environment (Figure \ref{figshear}).

\section {Supernovae rates}
\begin{figure}
\includegraphics{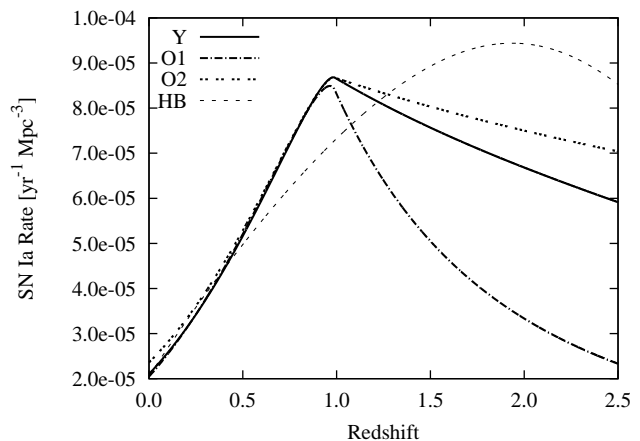}
\caption{The rates of non-lensed supernovae type Ia for different SFH function (from \citet{2011MNRAS.417..916G}).}
\label{figIarate}
\end{figure}
\begin{figure}
\includegraphics{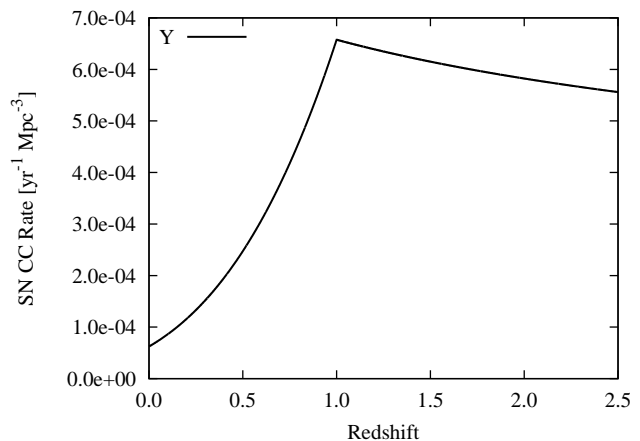}
\caption{The rates of non-lensed CC supernovae for SFH function Y only (from \citet{2011MNRAS.417..916G}).}
\label{figCCrate}
\end{figure}

The volumetric supernova type Ia rate at the cosmic time t - $R_{Ia}(t)$ is the convolution of the cosmic star formation history (SFH) $S(t)$ with the delay-time distribution (DTD) $\Psi(t)$:
\begin{eqnarray}
R_{\mathrm{Ia}}(t)=\int_0^t S(t-\tau)\Psi(\tau)d\tau.
\end{eqnarray}
The SFH function has been measured out to redshift 8 (\citealt{2008ApJ...683L...5Y}).
Following \citet{2011MNRAS.417..916G} we use four different SFHs in order to investigate any dependance of  lensed supernovae rates on the SFH function.
The SFH functions are defined as broken power laws in the form of $S(z)\sim(1+z)^{\gamma}$, as proposed by \citet{2008ApJ...683L...5Y} (hereafter denoted as Y) and \citet{2008PASJ...60..169O} (hereafter O1 and O2) and $S(z)\sim(a+bz)/(1+(z/c)^d)$, as defined by \citet{2006ApJ...651..142H} (HB).
\citet{2011MNRAS.417..916G} show that only power-laws of the form $\Psi(t)\sim(t/1Gyr)^{-\beta}$ for the DTD function are consistent with the data.
Table \ref{TabSFH} contains best-fitted values of all parameters from \citet{2011MNRAS.417..916G} for all SFH definitions used here.
Different SN Ia rates resulting from different SFHs are presented in Figure \ref{figIarate}.

\begin{figure}
\includegraphics{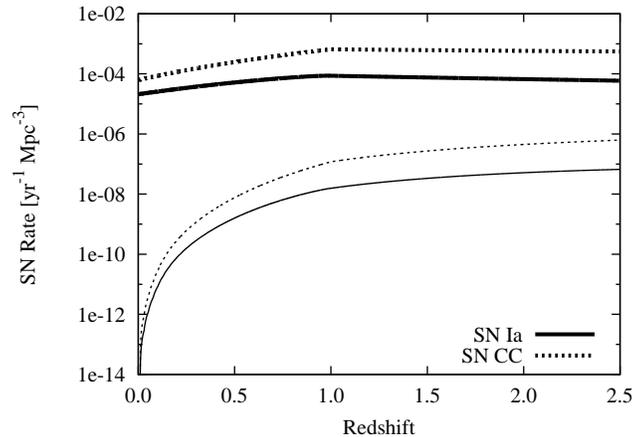}
\caption{The rates of lensed (lower curves) and non-lensed (upper curves) supernovae of SNe Ia (solid lines) and SNe CC (dashed lines). The rates were calculated supposing the SFH Y function.}
\label{RateAllLensed}
\end{figure}

For the core-collapse supernovae we also use a rescaled SFH function because of the suggestion that this rate is simply proportional to the cosmic star formation rate.
In Figure \ref{figCCrate} we plot the core-collapse SN rate with the normalisation from \citet{2011MNRAS.417..916G}.
Moreover, in Figure \ref{RateAllLensed} we show the comparison between the rates of lensed and non-lensed supernovae of both types.

\begin{table}
\centering
\caption{Star formation histories and results of best-fitting DTD parameters.}
\begin{tabular}{c l c}
\hline
SFH & Parametrization SFH & DTD parameter $\beta$ \\
\hline
Y & $\gamma_1=3.4, z_b=1, \gamma_2=-0.3$ & 1.1 \\
O1 & $\gamma_1=3.0, z_b=1, \gamma_2=-2.0$ & 1.23 \\
O2 & $\gamma_1=4.0, z_b=1, \gamma_2=0.0$ & 0.96 \\
HB & $a=0.0170, b=0.13, c=3.3, d=5.3$ & 1.11 \\
\hline
\end{tabular}
\label{TabSFH}
\end{table}

\section {Lensed supernovae}
\begin{figure}
\includegraphics[scale=0.65]{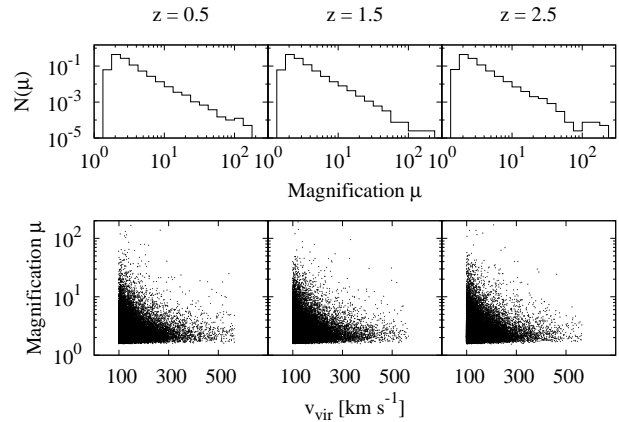}
\caption{Upper panel: The magnification distribution for sources at redshifts 0.5, 1.5, 2.5. Lower panel: The magnification versus lens virial velocity (used for estimating the mass) for sources located at redshifts 0.5, 1.5, 2.5.}
\label{figmag}
\end{figure}
\begin{figure}
\includegraphics{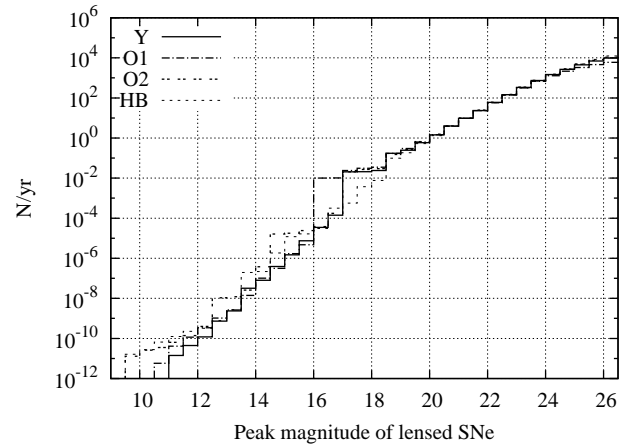}
\caption{The cumulative number of lensed SNe Ia per year per whole sky (peak magnitudes in rest-frame B-band) for different SFH.}
\label{LrateIa}
\end{figure}
\begin{figure}
\includegraphics{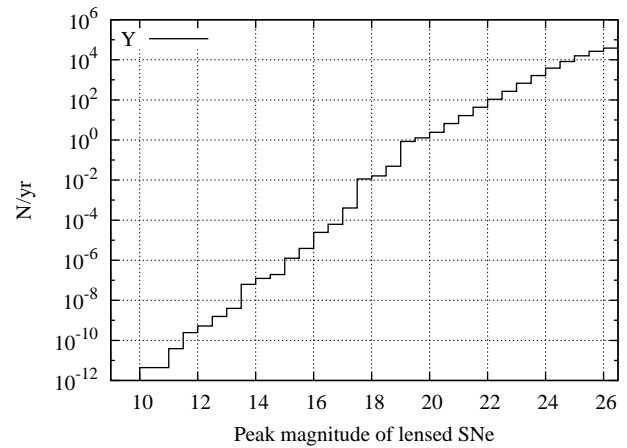}
\caption{The cumulative number of lensed SNe CC per year per whole sky (peak magnitudes in rest-frame B-band).}
\label{LrateCC}
\end{figure}

We choose the singular isothermal ellipsoid (SIE) model with external shear as the lens model for haloes from the Millennium Simulation. 
The SIE model is parametrized by the angular Einstein radius $\Theta_{\mathrm{SIE}}$, which is related to the virial velocity $v_{vir}$ of the lens galaxy that we randomise based on the distribution obtained from the Millennium Simulation. 
We assume a Gaussian distribution for the axis ratio with a mean of 0.7 and dispersion of 0.16 as in \citet{2010MNRAS.405.2579O}. The distribution is truncated at $q=0.5$ and $q=1$. 

Using simulated data we model about 350,000 lensing events of SNe sources in the \textsc{gravlens} programme \citep{2001astro.ph..2340K}.
For each system we obtain the image positions and the image magnifications of sources randomly situated in the background of lenses.
The position of the source (x,y relative to the lens axis) fulfils the condition $x^2+y^2<\Theta_{\mathrm{SIE}}^2$. 
By choosing the image with the highest magnification we can calculate the brightest observed magnitude of this lensed supernova.
It should be stressed that no a priori assumptions were made about the resulting images e.g. magnification, separation etc.
However, because the source is located inside the Einstein radius, there is more than one lensed image in the strong lensing case.
Our research concerned the lensing statistic, hence we are content with detecting even just one image at a given magnitude threshold.
From this simulation we obtain the amplification distribution for sources at different redshifts, see upper panel in Figure \ref{figmag}.
As expected, the most probable magnifications are near 2, but there is a non-negligible probability of obtaining magnifications in excess of 10 or 100.
The highest magnifications are due to specific source-lens system parameters, especially the position of the source, but not due to the source redshift, what was investigated in three separate redshift bins, shown in Figure \ref{figmag}. 
Lower panel of this figure shows that a source can be highly magnified even by a dark matter halo with non-high velocity (and therefore mass), which can occur at all redshifts.

Supernovae type Ia are thought to have the same value of absolute magnitude $M_B$ at the maximum of the light curve in B-band. 
This fact is ascribed to the origin of these objects as we mentioned in introduction.
However, the same cannot be said of core-collapse supernovae.  
Hence, we assume the average absolute magnitude of a supernova type Ia as $M_B=-19.13$ mag (\citealt{2011ApJS..192....1C}; \citealt{2011ApJ...737..102S}). 

Core-collapse SNe have a wide range of absolute magnitudes and we choose $M_B=-18.5$ mag, as an average value based on SN light curve templates from \citet{2002PASP..114..803N} and \citet{2012ApJ...745...32B}. 
The lensed supernovae will be magnified by a factor {\bf $-2.5\log(\mu)$} where in our case $\mu$ is the largest magnification in the simulated lens system.
The peak of the observed light curve peak has rest-frame B-band magnitude of 
\begin{eqnarray}
m_B^{peak}=M_B+5\log\left(\frac{d_L(z)}{10 \mathrm{pc}}\right)-2.5\log(\mu),
\end{eqnarray}
where $d_L(z)$ is the luminosity distance.

Combining the SN rates and the magnification distribution we obtain lensed supernovae rates for both types: Ia and core-collapse. 
First of all, Figure \ref{LrateIa} shows that we expect about 0.6 (between 0.53 and 0.66 for different SFH functions) lensed supernova type Ia per year up to magnitude 20 on the whole sky. 
However, for future surveys (with 24 magnitude capability) we predict about 700 detectable SNe type Ia per year on the whole sky.

Nonetheless, a closer look at the rates histogram reveals that those measurements cannot be expected to easily distinguish between different SFH functions. 
As seen in Fig. \ref{figIarate} SFH functions differ significantly only at higher redshifts, i.e. above 1.5.
As most SNe are magnified only by a small factor these redshifts correspond mostly to SNe of magnitude above 24.5, which is near the limit of our study.

For the core-collapse SNe we predict 1.3 and less than 1700 lensed objects per year for the whole sky within 20 and 24 mag range, respectively (Figure \ref{LrateCC}). 

In this study we focus on detectability of the strongest magnified image of a supernova. 
However, when the separation between the lensed images is smaller than the angular resolution of a detector, the magnified images will merge, forming a single blended observed image. 
We tried to estimate the brightness of such blended images taking into account both temporal evolution of the light curve of the lensed supernova and the time delay between the images. In our sample of lenses most image separations fall below 0.5 arc seconds for double-imaging lenses and below 0.2 arc seconds for quads, which is comparable with the resolving powers of typical ground-based surveys.
For the two-image systems (more than 97 per cent of systems in whole sample), only in case of 8 per cent the shift in observable brightness is significantly higher than in case of a single image by at least of 0.5 mag. 
The average shift in brightness is $0.23\pm0.17$ mag, which is comparable with the error in the estimation of the absolute magnitude for the SNe of both types.
In case of the four-image systems the overall brightness of blended images can be much larger than the brightness of the strongest image. 
The average shift in brightness is about 1 mag, reaching up to 1.4 mag, however, the quads compose only 3 per cent of our sample, hence the overall effect of blending of the images can be neglected.

We do not take into account microlensing of lensed SNe caused by the stars of the lens galaxy \citep{2006ApJ...653.1391D} as this would be beyond the scope of this work. 
We also neglect the effect of extinction in the foreground galaxy, which is claimed to affect neither the detectability of lensed events nor the flux measurements \citep{2002A&A...393...25G}.

Owing to the fact that we are interested in the statistical probability of lensed supernovae appearance we do not investigate the impact of the source population characteristic.
Obviously this seems to be important during choosing specific lens systems for observation.

\begin{table*}
\centering
\caption{The expected number of lensed SNe type Ia and core-collapse SNe in various surveys (expected efficiency about 100 and less than 20 per cent).}
\begin{tabular}{c c c c c c c c}
\hline
Survey & Survey area & Survey time & Survey limiting & Number SNIa & Number SNCC & Number SNIa & Number SNCC \\
& $\mathrm{[deg^2]}$ & [years] & magnitude (filter) & [1 year] & [1 year] & [survey time] & [survey time] \\
\hline
Stripe 82 (100\%) & 300 & 8 & 21.5 (i) & 0.073 & 0.124 & 0.584 & 0.992 \\
Gaia (100\%) & 40000 & 5 & 20.0 (V) & 0.60 & 1.27 & 3.0 & 6.35 \\
LSST (100\%) & 20000 & 10 & 24.5 (i) & 710. & 1900. & 7100. & 19000. \\
\hline
Stripe 82 (20\%) & 300 & 8 & 21.5 (i) & 0.015 & 0.025 & 0.119 & 0.198 \\
Gaia (20\%) & 40000 & 5 & 20.0 (V) & 0.12 & 0.254 & 0.6 & 1.27 \\
LSST (20\%) & 20000 & 10 & 24.5 (i) & 142. & 380. & 1420. & 3800. \\
\hline
\end{tabular}
\label{TabSurvey}
\end{table*}

\section{Discussion \& Conclusions}

\begin{figure*}
\includegraphics[scale=1.40]{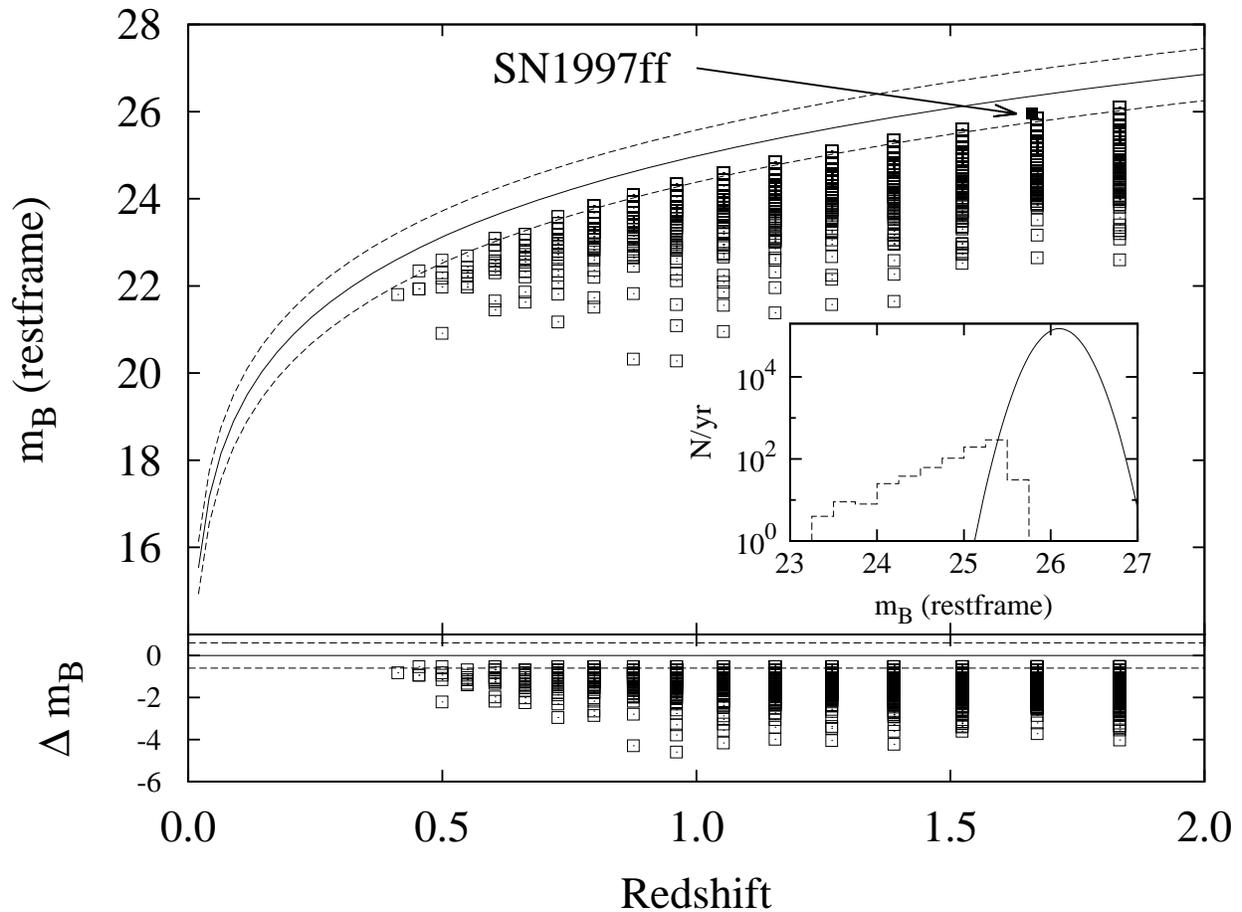}
\caption{The Hubble diagram for lensed SNe. The solid line represents distance modulus for cosmological parameters: $\Omega_{\mathrm{M}}=0.27$, $\Omega_{\Lambda}=0.73$, dotted line denotes the scatter in current combined sample of SNe Ia \citep{2011ApJS..192....1C}, boxes - simulated lensed SNe Ia for 1-year-observations on the whole sky with 100 per cent efficiency. Lower panel - residuals from the cosmological model.
Inset: distribution of number of lensed and non-lensed SNe Ia in the sample at redshift of about 1.5. The solid line represents the non-lensed sample, the dashed line - the lensed sample.
The arrow points to location of a probable lensed supernova SN1997ff \citep{2001MNRAS.324L..25L}.}
\label{Hubb}
\end{figure*}

In this paper we present the simulations of supernovae being lensed by galaxies, represented by haloes from Millennium Simulation. 
The numbers of expected lensed supernovae in large scale surveys are gathered in Table \ref{TabSurvey} for SDSS Stripe82, Gaia and LSST. 
Given their limiting magnitude, assuming about 20 per cent detection efficiency we estimate that in the entire Stripe 82 data (about 8 years) there is hardly any chance (10 and 20 per cent for SN Ia and CC, respectively) that the data contains any lensed supernova, unless we got really lucky. 

Gaia mission will cover the entire sky in its 5 years (2013-2018) regular observations. 
Its sampling will depend on the region of the sky, hence the efficiency will vary. 
Assuming 20 per cent average detection efficiency we estimate that during the entire mission there might be one lensed supernova detected of each type. 
Gaia data processing pipeline will have the special ability to classify supernovae into types and to estimate the redshifts based on the low-resolution spectrograph available on-board (\citealt{2012IAUS..285..425W}, \citealt{2012arXiv1210.5007W}, Blagorodnova et al. in prep.). 
Therefore, a lensed case of supernova should be recognised promptly and will be alerted to the community immediately. 

Because of its much deeper imaging capabilities, the LSST (planned to operate from 2018-2020) will give us more rewarding results with a significant number of detections of lensed supernovae. 
Again, assuming 20 per cent efficiency for the part of the sky LSST will cover in 10 years of expected operation there should be about 1400 lensed supernovae type Ia and more than 3800 lensed core-collapse supernovae. 
This will significantly enhance our probing of the cosmological distance scale with large statistical samples of high redshift supernovae.
The estimated number of SN detections are higher than the estimates of \cite{2010MNRAS.405.2579O} on account of no assumptions being made on detection of a second image (e.g. separation and magnitude limit), which being essential in that study.

Our simulation indicates that sources at redshift larger than 0.5 have lensing probability better than $10^{-5}$ (Figure \ref{figlike}) and vast majority of lensed supernovae originate from these redshifts. 
The rates of lensed supernovae of redshift 0.5 and higher are about $10^{-9}$ to $10^{-7}$, indicating these are still relatively rare events, even in high redshifts. 
When compared to the rates of all supernovae we expect that at redshifts 0.5 and higher about 0.06 per cent of all supernovae will be lensed at least by a factor 2 (Figure \ref{RateAllLensed}). 
The majority of the brightest images of lensed SNe will be magnified by factor 2 (i.e. by 0.75 mag), however not negligible part of them will have magnifications exceeding 10, making the lensed supernovae easily standing out of the bulk of typical supernovae.

The hunt for lensed supernovae can be also performed in a targeted regular observations of selected known galaxy-galaxy lenses. 
Our study demonstrate that the most important factor boosting the possibility of higher magnification is the position of the lensed galaxy on the source plane. 
If the supernova host galaxy intersects caustics we will obtain higher probability of larger magnification, especially if we expect more supernovae to burst closer to the centre of the source galaxy.

Moreover, when looking for supernovae lensed from redshift 0.5, their B-band peak at the rest-frame due to cosmological shifting will occur in near-infra red. 
Therefore, observing at these wavelengths will assure the supernova is observed at its brightest, hence improving chances of its detection.

We also investigate the impact of lensed supernovae type Ia on the Hubble diagram (Fig. \ref{Hubb}).
Generally, it can be claimed that the majority of lensed supernovae sample will be distinguishable from the current sample. 
However, every outlier should be checked against being contaminated by a lensing event.
The inset in Figure \ref{Hubb} shows whether two samples - non-lensed and lensed supernovae type Ia which burst in redshift bin between 1.4 and 1.5 - are distinguishable.
Here we assume average SN magnitude as 26.1 with dispersion $\sim$0.2 \citep{2011ApJS..192....1C}.
The dispersion is the result of an intrinsic scatter in observations due to an incomplete understanding of SN physics or all photometric uncertainties.
It can be expected that a part of a lensed sample is mixed with an non-lensed sample. 
For example, SN 1997ff discovered in the {\it Hubble Deep Field} was identified as a high-redshift supernova mildly magnified by gravitational lensing \citep{2001MNRAS.324L..25L}.
Two galaxies lying close to the line-of-sight enhanced the magnitude of this SN by $\sim$0.4 mag (as indicated in Fig. \ref{Hubb}), a change larger than the differences of various cosmological models' predictions.

Our results are in strong agreement with a recent study by \cite{2012arXiv1207.3708K}, where they demonstrate a Bayesian statistical methodology for constraining the properties of dark matter haloes of foreground galaxies that intersect the lines-of-sight towards supernovae type Ia. 
The method was applied to realistic simulated SNIa data, based on the real 3-year Supernova Legacy Survey (SNLS3) data release (about 160 SNeIa at redshifts $0.1<z<1$ up to 24.5 mag, on 4 deg$^2$ survey area). 
Two dark matter halo density profile of foreground galaxies models were chosen: a truncated singular isothermal sphere and a Navarro-Frenk-White profile.  
They show that detection of a lensing signal in the simulated sample is highly unlikely for both dark matter halo profiles. 
Albeit, in the analysis of real SNLS3 data they obtain a very minor detection of the lensing signal (for the truncated singular isothermal sphere model only).
For such a small survey area we expect as few as 0.028 lensed event per year (assuming 20 per cent detection efficiency). 

Our study shows that even though some surveys will detect large samples of lensed supernovae, the differences due to different SFH functions start to emerge only at a level of about 26 magnitude, not available yet in wide field surveys.  
However, the robust sample of lensed supernovae can give answers to a number of questions generated by the cosmological simulations results, e.g. about the universality of the dark matter density profiles.

\section*{Acknowledgments}
We are greatful to the Anonymous Referee, whose critical remarks greatly improved the paper.
We thank Dr Vasily Belokurov for initiating the idea of this research and Drs Phil Marshall and Matt Auger for useful comments.
We also thank Dr Dovi Poznanski and Or Graur for kindly and quickly providing information on SN rates.
The Millennium Simulation databases used in this paper and the web application providing online access to them were constructed as part of the German Astrophysical Observatory. 
This work has been supported in part by the Polish National Science Centre grant N~N203~581540.

\bibliographystyle{mn2e}

\label{lastpage}

\end{document}